\documentclass[useAMS,usenatbib]{mnras}

\usepackage{graphicx, bm, amssymb, xcolor}
\usepackage{verbatim}
\usepackage[all]{hypcap}

\usepackage{epsfig}
\usepackage{aas_macros}
\usepackage{natbib}
\usepackage{times}

\newcommand{\msun}{\mbox{M$_\odot$}}
\newcommand{\yr}{\mbox{${\rm yr}$}}
\newcommand{\myr}{\mbox{${\rm Myr}$}}

\newcommand{\pc}{\mbox{${\rm pc}$}}
\newcommand{\kpc}{\mbox{${\rm kpc}$}}

\newcommand{\myk}{\mbox{$\msun~\yr^{-1}~\kpc^{-2}$}}
\newcommand{\be}{\begin{equation}}
\newcommand{\ee}{\end{equation}}
\newcommand{\bea}{\begin{eqnarray}}
\newcommand{\eea}{\end{eqnarray}}

\def\fbound{\hbox{$f_{\rm bound}$}}
\def\fass{\hbox{$f_{\rm assoc}$}}

\defcitealias{kruijssen12d}{K12}
\defcitealias{chandar15}{C15}

\title{Pitfalls when observationally characterizing the relative formation rates of stars and stellar clusters in galaxies}
\author{J.~M.~Diederik Kruijssen$^{1}$\thanks{kruijssen@uni-heidelberg.de} and Nate Bastian$^2$\thanks{n.j.bastian@ljmu.ac.uk}\\
$^1$Astronomisches Rechen-Institut, Zentrum f\"{u}r Astronomie der Universit\"{a}t Heidelberg, M\"{o}nchhofstra\ss e 12-14, 69120 Heidelberg, Germany\\
$^2$Astrophysics Research Institute, Liverpool John Moores University, IC2, Liverpool Science Park, 146 Brownlow Hill,
Liverpool L3 5RF, United Kingdom}

\begin{document}

\date{Accepted 2015 November 10. Received 2015 November 2; in original form 2015 September 16.}

\pagerange{\pageref{firstpage}--\pageref{lastpage}} \pubyear{2015}

\maketitle

\label{firstpage}

\begin{abstract}
Stars generally form in aggregates, some of which are bound (`clusters') while others are unbound and disperse on short ($\sim10~\myr$) timescales (`associations'). The fraction of stars forming in bound clusters ($\Gamma$) is a fundamental outcome of the star formation process. Recent observational and theoretical work has suggested that $\Gamma$ increases with the gas surface density ($\Sigma$) or star formation rate (SFR) surface density ($\Sigma_{\rm SFR}$), both within galaxies and between different ones. However, a recent paper by Chandar et al.~has challenged these results, showing that the {\it total} number of stellar aggregates per unit SFR does not vary systematically with the host galaxy's absolute SFR. In this Letter, we show that no variations are expected when no distinction is made between bound and unbound aggregates, because the sum of these two fractions should be close to unity. We also demonstrate that any scaling of $\Gamma$ with the absolute SFR is much weaker than with $\Sigma_{\rm SFR}$, due to the mass-radius-SFR relation of star-forming `main sequence' galaxies. The environmental variation of $\Gamma$ should therefore be probed as a function of area-normalised quantities, such as $\Sigma$ or $\Sigma_{\rm SFR}$. We present a set of guidelines for meaningful observational tests of cluster formation theories and show that these resolve the reported discrepancy.
\end{abstract}


\section{Star formation in bound stellar clusters and in unbound associations}
\label{sec:bound}
Stars appear to rarely form in isolation, with the majority forming as parts of stellar aggregates. These aggregates vary strongly in their (surface) densities, with some forming in compact, gravitationally bound groups that may survive for tens of Myr or more (generally referred to as ``clusters") and others forming looser systems (associations) that are likely unbound \citep[e.g.][]{bressert10,gieles11}. Whether the fraction of stars forming in either bound clusters (\fbound, also referred to as $\Gamma$, \citealt{bastian08}) or associations (\fass) varies as a function of environment has been a long standing issue within the community. Irrespective of the answer to this question, it is generally agreed that $\fbound + \fass \approx 1$ (and hence $\fass\approx1-\fbound=1-\Gamma$), as few stars appear to be formed in isolation, or in systems that would not be considered clusters or associations \citep{dewit05}.

In order to determine the fraction of stars that form in bound clusters, recent observational studies have attempted to separate clusters from associations by either using high resolution HST imaging to select centrally concentrated and roughly spherical objects (indicative of having evolved over multiple dynamical timescales, see \citealt{gieles11}) or by focussing on slightly older systems ($10-50$~Myr) when associations have already dispersed into the field. The general result of these studies is that galaxies (and regions within galaxies) with high gas and star formation rate (SFR) densities ($\Sigma$ and $\Sigma_{\rm SFR}$) tend to form a higher fraction of their stars in bound clusters (see the compilation by \citealt{adamo15b}).

These observational results have been complemented by theoretical developments on how star and cluster formation are influenced by the properties of the interstellar medium (ISM) from which stars form \citep[hereafter \citetalias{kruijssen12d}]{kruijssen12d}. In the \citetalias{kruijssen12d} model, star formation takes place across the density spectrum of the ISM. However, only the highest density peaks have free-fall times short enough to reach the high star formation efficiencies necessary for stellar aggregates to remain bound upon residual gas expulsion. Higher-pressure ($P$) galaxies (observationally traced by higher $\Sigma\propto P^{0.5}$ and $\Sigma_{\rm SFR}\propto P^{0.7}$, see e.g.~\citealt{kennicutt98b,krumholz05}) have an ISM characterised by higher-density peaks and therefore higher star formation efficiencies, resulting in a larger fraction of stars forming in bound clusters. The basic prediction is that $\Gamma$ increases with $P$, $\Sigma$, and $\Sigma_{\rm SFR}$, from $\Gamma\sim0.01$ at $\Sigma_{\rm SFR}=10^{-3}~\myk$ to $\Gamma\sim0.5$ at $\Sigma_{\rm SFR}=10^{0}~\myk$, in quantitative agreement with the observational results. The physics contained in the \citetalias{kruijssen12d} cluster formation model also form the basis of now-standard models for galactic-scale star formation, which reproduce the observed relations between the gas mass and the SFR \citep[e.g.][]{krumholz05,padoan11,hennebelle11,federrath12}.

Recently, \citet[hereafter \citetalias{chandar15}]{chandar15} have reported observations that they interpret as being in contradiction to the above observational and theoretical studies. The authors analyze the mass functions of stellar aggregates in seven star-forming galaxies and normalise them by the derived galaxy-wide SFRs. In principle, if more stars were being formed in aggregates (per unit SFR), these normalised mass functions should be higher than if less stars were being formed in aggregates. \citetalias{chandar15} find little such dependence on the SFR.

However, the authors make no distinction between bound or unbound aggregates, therefore their sample consists of the sum of all young ($<10$~Myr) stellar aggregates (i.e.~$\fbound + \fass\approx1$). As such, we would not expect to see much variation, as the only expected variation would come from the fraction of isolated star formation or star formation that would not be characterised as a cluster or association. In principle, this could be verified explicitly for the \citetalias{chandar15} catalogues by comparing the cluster formation rates to the total SFRs, if the catalogues were publicly available. However, the cluster catalogues for most of the galaxies considered in their study have not been made public. In the one case where the catalogue has been shared (for the grand-design spiral galaxy M83), it was shown by \citet{bastian12} that for ages $\tau<10~\myr$, the catalogue is dominated by irregular, low-density aggregates reminiscent of the unbound associations observed in the solar neighbourhood. While this result suggests that the cluster catalogues of the other galaxies in \citetalias{chandar15} are likely also dominated by low-density aggregates at young ages, this cannot be verified. Due to this uncertainty, we leave the fraction of unbound associations that is included in the sample as a free parameter throughout the rest of this work. The effect of including large fractions of unbound associations within the analysis is quantified below in \S\ref{sec:plot}.

\citetalias{chandar15} also include the results derived from old aggregates ($100$--$400$~Myr) within these seven galaxies. Due to their age \citep[and number of crossing times that they have undergone,][]{gieles11} these systems are almost certainly all stellar clusters. However, there are a number of important caveats in using such old systems. The first is that it implicitly assumes that the SFR observed today (as measured through H$\alpha$ emission) is representative of the SFR of the systems $100$--$400$~Myr ago, which may be adequate for quiescent galaxies, but is unlikely to be true in ongoing starbursts in the Antennae or the central regions of M83. The second is that the population may be affected by cluster disruption, with some galaxies losing large fractions of the clusters by this age (e.g., the Antennae or central regions of M83) while other galaxies will not have had their cluster populations strongly affected \citep[e.g.][]{annibali11,kruijssen11,kruijssen12c,baumgardt13,adamo15}. Because the disruption of gravitationally bound clusters proceeds on shorter timescales in higher-density environments, the use of older clusters washes out the increase of $\Gamma$ with the (surface) density or pressure \citep[e.g.][]{kruijssen11,bastian12}.

\section{The fraction of stars forming in bound clusters as a function of the host galaxy properties}
\label{sec:galaxy}
\citetalias{chandar15} compare their observational results to the \citetalias{kruijssen12d} cluster formation theory, concluding that their observations are not in agreement with the model predictions. Unfortunately, the \citetalias{kruijssen12d} model was not correctly compared to the observations, meaning that the conclusions reached by the authors are not valid. The first reason why the comparison does not hold was already given in \S\ref{sec:bound} -- \citetalias{chandar15} consider all stellar aggregates, regardless of whether they are bound clusters or unbound associations, whereas the \citetalias{kruijssen12d} model predicts the fraction of stars forming in gravitationally bound clusters only.

There is also a second reason why the comparison of the observations to the model is not adequate. In the \citetalias{kruijssen12d} model, the relevant physical parameter that controls $\Gamma$ is not the absolute SFR, but rather the SFR surface density $\Sigma_{\rm SFR}$, which is intricately linked to the state of the ISM of the host galaxy (specifically the gas pressure, as traced by the gas surface density). By contrast, \citetalias{chandar15} only use the absolute SFR. In order to convert the SFR to $\Sigma_{\rm SFR}$, one needs to assume a relation between the galaxy radius and its SFR. The exact relation used by \citetalias{chandar15} is not given. However, the \citetalias{kruijssen12d} model in Figure~4 of \citetalias{chandar15} scales linearly with the SFR, which is the same slope as the steepest scaling with $\Sigma_{\rm SFR}$ for $\Sigma_{\rm SFR}<10^{-2}~\myk$ (see Figure~6 and Equation (45) of \citetalias{kruijssen12d}). We thus infer that a constant radius for all galaxies must have been assumed to convert the model dependence on $\Sigma_{\rm SFR}$ to a dependence on the SFR. This assumption is clearly unphysical, yet it greatly impacts the results.

In late-type, star-forming galaxies, the radius scales with stellar mass as a power law $R \propto M^{\alpha}$ with $\alpha=0.14-0.39$ \citep{shen03}. For galaxies on the main sequence, the stellar mass scales near-linearly with the SFR as ${\rm SFR}\propto M^{0.9}$ \citep{brinchmann04,peng10}. We can therefore write $R\propto {\rm SFR}^{\beta}$ with $\beta=0.16-0.43$. This means that the SFR scales with the SFR density as $\Sigma_{\rm SFR} \propto {\rm SFR}^{1-2\beta} \propto {\rm SFR}^{\eta}$ with $\eta=0.13-0.69$ and a mean of $\eta\sim0.4$.

While the exact slope of the $\Sigma_{\rm SFR}$-${\rm SFR}$ relation is uncertain, the critical point is that the slope of this relation greatly affects the prediction in the ${\rm SFR}$-$\Gamma$ plane. If a constant radius is assumed, then $\Sigma_{\rm SFR} \propto {\rm SFR}$ instead of $\Sigma_{\rm SFR} \propto {\rm SFR}^{0.4}$ (taking the mean of the range of $\eta$ given above). Given the large scatter on the main sequence of galaxies as well as the mass-radius relation, neither transformation is necessarily recommended. However, $\Sigma_{\rm SFR} \propto {\rm SFR}^{0.4}$ is more accurate than assuming a single radius for all galaxies independently of their mass -- especially given that the \citetalias{chandar15} galaxy sample spans almost 3 orders of magnitude in mass and SFR. If the \citetalias{kruijssen12d} model predicts $\Gamma \propto \Sigma_{\rm SFR}^{\zeta}$ (with some value for $\zeta$), then the assumption of a constant radius will imply $\Gamma \propto {\rm SFR}^{\zeta}$, whereas using the mass-radius relation of star-forming galaxies on the main sequence implies $\Gamma \propto {\rm SFR}^{0.4\zeta}$. \citetalias{chandar15} thus incorrectly apply the \citetalias{kruijssen12d} model by artificially boosting its intrinsically-shallow increase with the SFR by a factor of 2.5.

\section{Pitfalls when addressing the fraction of stars forming in bound clusters}
\label{sec:plot}
\begin{figure}
\includegraphics[width=\hsize, angle=0]{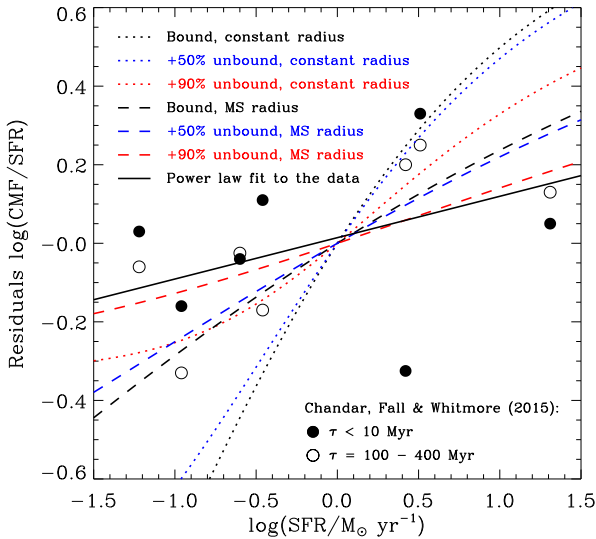}
\caption{Relative change of the cluster mass function (CMF) normalised to the galaxy-wide SFR as a function of the SFR. Symbols show the seven galaxies from \citetalias{chandar15} for stellar aggregates with ages $\tau\leq10~\myr$ (filled) and $\tau=100$--$400~\myr$ (open). The black solid line shows a power-law fit to all data points. The dotted lines show the prediction of the \citetalias{kruijssen12d} model when converting its dependence on $\Sigma_{\rm SFR}$ to a dependence on the SFR using a constant radius for all galaxies. The dashed lines show the \citetalias{kruijssen12d} model when adopting the observed relation between the $\Sigma_{\rm SFR}$ and the SFR for star-forming main sequence galaxies. The black dotted and dashed lines only consider the normalised mass function of bound clusters, whereas the blue and red lines also include 50\% and 90\% of the unbound regions, respectively. When matching the assumptions made in the analysis by \citetalias{chandar15} (red dashed line), the model provides a clear match to the observations (black solid line).}
\label{fig:fig}
\end{figure}
In Figure~\ref{fig:fig}, we demonstrate how the assumptions made by \citetalias{chandar15} quantitatively affect the change of normalization of the SFR-scaled cluster mass function, ${\rm CMF}/{\rm SFR}$. \citetalias{chandar15} assume that this quantity scales linearly with $\Gamma$, but this is only true if exclusively bound clusters are considered. The dotted and dashed black lines in Figure~\ref{fig:fig} indicate the predictions of \citetalias{kruijssen12d} for the fraction of stars forming in bound stellar clusters $\Gamma$ as a function of the SFR.\footnote{In principle, the black dotted line in Figure~\ref{fig:fig} should be the same as the dotted line in Figure~4 of \citetalias{chandar15}. However, their line has a constant slope, whereas the actual prediction of the \citetalias{kruijssen12d} is curved as in our Figure~\ref{fig:fig}. The model only has a constant slope for $\Sigma_{\rm SFR}<10^{-2}~\myk$ and flattens for higher SFR surface densities, implying that \citetalias{chandar15} incorrectly approximate the model with a single slope across the full range of $\Sigma_{\rm SFR}$.} Relative to these lines, we see that including 50\% (blue lines, $f_{\rm incl}=0.5$) or 90\% (red lines, $f_{\rm incl}=0.9$) of the unbound regions (i.e.~associations) strongly weakens the predicted increase with the SFR, because ${\rm CMF}/{\rm SFR}\propto \fbound+f_{\rm incl}\fass \approx \Gamma + f_{\rm incl}(1-\Gamma)$. The lines show that the change between $f_{\rm incl}=0$ and $f_{\rm incl}=0.5$ is modest, but the model dependence on the SFR strongly flattens when including more than 50\% of the unbound regions. If 100\% of the unbound regions are included (i.e.~$f_{\rm incl}=1$), the model line is flat by definition, because the total number of aggregates (${\rm CMF}/{\rm SFR}\propto \fbound + \fass \approx 1$) is then covered at all SFRs. These examples show that the \citetalias{chandar15} results are consistent with the prediction of \citetalias{kruijssen12d} when the inclusion of unbound aggregates is accounted for.

Figure~\ref{fig:fig} also quantifies how the theoretically-predicted variation of ${\rm CMF}/{\rm SFR}$ with the SFR is affected by the adopted galaxy radius. The dotted lines show the prediction of \citetalias{kruijssen12d} when translating the predicted $\Sigma_{\rm SFR}$-$\Gamma$ relation to a dependence on ${\rm SFR}$ by using a constant galaxy radius as in \citetalias{chandar15}, which is inconsistent with theory and observations. By contrast, the dashed lines show the model predictions when adopting the observed relations between the radii and SFRs of star-forming main sequence galaxies. Similarly to the effect of including unbound associations, we see that the model prediction is strongly flattened by adopting the scaling between radius and SFR that is actually appropriate for the observations under consideration. This change is so pronounced that by itself it reconciles model and observations to within the observational scatter. Indeed, a comparison of the black dashed and dotted lines in Figure~\ref{fig:fig} shows that \citetalias{chandar15}'s choice of a constant galaxy radius affects the ${\rm CMF}/{\rm SFR}$-${\rm SFR}$ relation even more strongly than the inclusion of 90\% of all unbound associations (cf.~the red and black dotted lines).

The red dashed line in Figure~\ref{fig:fig} makes the same assumptions as the observations presented in \citetalias{chandar15}, by including all bound stellar clusters and 90\% of the unbound associations, as well as adopting the observed $R$-${\rm SFR}$ relation of star-forming galaxies. The power-law slope of this model (${\rm CMF}/{\rm SFR}\propto{\rm SFR}^{0.11}$) is identical to that of a power-law fit to all observations (black solid line) at the two-decimal precision adopted here, and closely resembles the slope of a power-law fit to the filled data points only (which represent aggregates with ages $\tau<10~\myr$ and have ${\rm CMF}/{\rm SFR}\propto{\rm SFR}^{0.04}$). This shows that theory and observations agree when the physically relevant comparison is made.

Likewise, we can exclusively consider the open data points in Figure~\ref{fig:fig}, which refer to the old aggregates with ages ($\tau=100$--$400~\myr$). These data points should be compared to the model that does not include unbound associations (which disperse on timescales $<100~\myr$) and uses the observed $R$-${\rm SFR}$ relation (black dashed line, with ${\rm CMF}/{\rm SFR}\propto{\rm SFR}^{0.26}$). Inspection of Figure~\ref{fig:fig} shows that this model indeed provides a good match to the observations, as a power-law fit to the open symbols satisfies ${\rm CMF}/{\rm SFR}\propto{\rm SFR}^{0.17}$. The small difference relative to the model is likely due to cluster disruption, which is not included in the model but is expected to increase in strength towards higher SFRs (see e.g.~Table~2 of \citealt{fall12} and \citealt{kruijssen11,kruijssen12c}), thereby dampening the increase with the SFR of the open data points in Figure~\ref{fig:fig}. The isolation of the older cluster populations (and their agreement with the black dashed rather than the red dashed model line) thus shows that not only are the observations reproduced by the model, but they also follow the predicted age dependence.

\section{A set of guidelines for testing cluster formation theories}
\label{sec:guide}
In summary, the model line shown in Figure~4 of \citetalias{chandar15} does not reflect the prediction of \citetalias{kruijssen12d}. The observations consider the combined sample of bound clusters and unbound associations, whereas the model concerns bound clusters only. In addition, the line of \citetalias{chandar15} translates the model to an observational plane (${\rm SFR}$-$\Gamma$) different than the plane the model was formulated in ($\Sigma_{\rm SFR}$-$\Gamma$), by applying a transformation from $\Sigma_{\rm SFR}$ to ${\rm SFR}$ that is inconsistent with observed galaxy properties. By correctly accounting for these two crucial factors, we demonstrate that the observations by \citetalias{chandar15} are fully consistent with the predictions of the \citetalias{kruijssen12d} model as well as with recent observations showing that the fraction of stars forming in bound clusters varies as a function of environment.

\begin{table}
 \centering
  \begin{minipage}{68mm}
  \caption{Galaxy properties}\label{tab:table}
  \begin{tabular}{@{}l c c r@{}}
  \hline
   Galaxy & $\log{(\Sigma/\msun~\pc^{-2})}$ & $\gamma$ & References$^a$ \\
  \hline
SMC & $0.96$ & $0.70$ & 1 \\
NGC 4214 & $0.96$ & $0.70^b$ & 2 \\
LMC & $1.04$ & $0.75$ & 1 \\
NGC 4449 & $1.08$ & $0.75^b$ & 3 \\
M51 & $1.47$ & $0.95^b$ & 1 \\
M83 & $1.70$ & $0.95$ & 1 \\
Antennae & $1.88$ & $1.05$ & 4 \\
  \hline
\end{tabular}\\
$^a$References: (1) \citetalias{kruijssen12d}, (2) \citet{leroy13}, (3) \citet{hunter99}, (4) \citet{zhang01}.\\
$^b$Estimated by adopting the value of the galaxy with the closest gas surface density.
\end{minipage}
\end{table}

Like all theories, the \citetalias{kruijssen12d} model has its limitations and its inevitable falsification will lead to key new insights into the cluster formation process. However, in order to achieve this goal, care must be taken to compare the right quantities. Whenever possible, the following guidelines should be followed.
\begin{enumerate}
\item
We recommend that future investigations of the environmental variation of $\Gamma$ only consider the relative contribution to the total SFR of {\it centrally-concentrated clusters}. A centrally-concentrated morphology is indicative (though not conclusive proof) of relaxation, which suggests survival over multiple dynamical times and thus separates bound clusters from unbound associations \citep{gieles11}. The obvious issue to bear in mind is that, for distant galaxies, even such catalogues will include some fraction of unbound associations due to the finite resolution of the observations. This uncertainty is largest for samples of aggregates with very young ages ($\tau<10~\myr$) and vanishes for $\tau>10~\myr$ \citep[Figure~6]{bastian12}. By contrast, cluster disruption is expected to affect the sample at ages $\tau>30~\myr$ \citep[e.g.][]{kruijssen11,bastian12}. We therefore recommend restricting the cluster sample to the age interval $\tau=10$--$30~\myr$ when measuring $\Gamma$ \citep[cf.][]{adamo15b}.
\item
In addition, future studies should determine $\Gamma$ as a function of the {\it area-normalised quantities} on which $\Gamma$ is predicted to depend, such as the SFR surface density, the gas surface density, or the gas pressure. The use of absolute quantities such as the total SFR or gas mass should be avoided, because these are simply proportional to the galaxy mass and provide little information on the {\it physical} conditions. For instance, galaxies with widely different masses and SFRs can have very similar gas or SFR surface densities. The dependence of the SFR on the galaxy mass thus inflates the dynamic range and gives the possibly misleading impression of covering a broad galaxy sample, even if the galaxies have very similar physical conditions. This is evident for the galaxies considered by \citetalias{chandar15}, which cover three orders of magnitude in SFR, but less than one order of magnitude in gas surface density.
\item
For each individual data point, the area for which these quantities are determined should be the same for the SFR, stellar clusters, and gas (if appropriate). There is no standard method for determining the star-forming area within galaxies. However, the scale-free nature of the \citetalias{kruijssen12d} model permits some freedom of choice. The area $A$ should be large enough to avoid the small-scale scatter caused by the time evolution of individual regions \citep[i.e.~$\sqrt{A/\pi}\ga0.5~\kpc$, see][]{kruijssen14}. In addition, the area should be small enough to not include significant empty space for any of the observables \citep[i.e.~$\sqrt{A/\pi}\la R_{25}$, where $R_{25}$ is the optical radius, see][]{kennicutt98b}.
\end{enumerate}

\begin{figure}
\includegraphics[width=\hsize, angle=0]{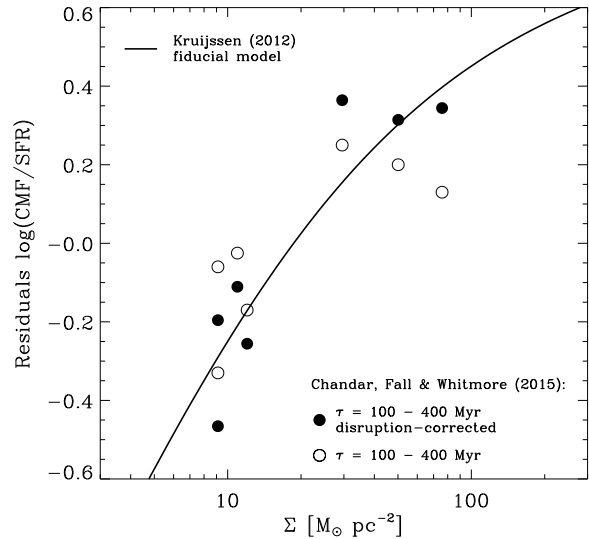}
\caption{Relative change of the cluster mass function (CMF) normalised to the galaxy-wide SFR as a function of the gas surface density $\Sigma$. Symbols show the seven galaxies from \citetalias{chandar15} for stellar aggregates with ages $\tau=100$--$400~\myr$, as observed (open) and corrected for cluster disruption (filled, which is a proxy for the relation at ages $\tau=10$--$40~\myr$; see the text for details). The black solid line shows the fundamental prediction of the \citetalias{kruijssen12d} cluster formation theory. The model line and the filled data points are in good agreement when the correct quantities are compared, following the guidelines provided in \S\ref{sec:guide}.}
\label{fig:fig2}
\end{figure}

\section{Implications of the presented guidelines}
\label{sec:concl}
We now demonstrate the application of these guidelines to the results of \citetalias{chandar15}. Fundamentally, the \citetalias{kruijssen12d} cluster formation theory predicts a dependence of $\Gamma$ on the gas surface density (or pressure).\footnote{The predicted dependence of $\Gamma$ on $\Sigma_{\rm SFR}$ is secondary, because it is only obtained by assuming a star formation relation between $\Sigma$ and $\Sigma_{\rm SFR}$ \citep[e.g.][]{kennicutt98b}.} Because the galaxies in the \citetalias{chandar15} sample are well-studied, gas surface densities are available for all of them and these are listed in Table~\ref{tab:table}. Figure~\ref{fig:fig2} now shows ${\rm CMF}/{\rm SFR}$ as a function of $\Sigma$. This addresses guideline (ii), because the gas surface density is an area-normalised quantity that reflects the physical conditions of star and cluster formation in each galaxy. We also address guideline (i), by only showing the intermediate-age clusters ($\tau=100$--$400~\myr$, open symbols) and thereby omitting unbound associations from the sample. However, as explained in \S\ref{sec:plot}, the cluster population is affected by disruption at these intermediate ages. The values of ${\rm CMF}/{\rm SFR}$ are decreased by disruption more strongly for the high-surface density points than for the low-surface density points, which decreases the slope of the open symbols. We therefore correct these points for cluster disruption based on the power law slopes $\gamma$ of the aggregate age distributions from \citet[where ${\rm d}N/{\rm d}\tau\propto\tau^{-\gamma}$]{fall12}. These slopes are also listed in Table~\ref{tab:table} and are used to estimate the residuals of ${\rm CMF}/{\rm SFR}$ at the age range $\tau=10$--$40~\myr$, as required by guideline (i).\footnote{For each galaxy, we have used the age distribution slope in the highest mass bin to ensure that our disruption estimates are conservative. \citet{fall12} only measured slopes for the SMC, LMC, M83 and Antennae galaxies. For the other three galaxies, we adopt the slope of the galaxy with the closest gas surface density (see Table~\ref{tab:table}).} In practice, this means that each data point is shifted up by $\gamma$ to estimate the number of clusters at ages exactly one order of magnitude younger than the original age range, after which the mean slope of the galaxy sample ($\gamma=0.84$) is again subtracted so that only the residuals are retained. Even though there are ways of accounting for cluster disruption that are much more sophisticated (e.g.~by using well-tested physical models), we prefer to use the aggregate age distributions from \citet{fall12} to minimise the possible points of contention.\footnote{We note that other works have found even shallower age distributions than \citet{fall12} for the low-density galaxies (SMC, LMC, NGC~4449) in the sample considered here \citep[see][]{adamo15}. Had we used these, it would have steepened the trend visible in Figure~\ref{fig:fig2} even further.} Finally, we note that guideline (iii) cannot be addressed for these galaxies, because the cluster catalogues are not publicly available and their spatial coverage is therefore unknown. We therefore assume that the coverage of the cluster catalogue in these galaxies encompasses most of the ongoing star formation.

The disruption-corrected values of ${\rm CMF}/{\rm SFR}$ resulting from the above procedure are shown as filled symbols in Figure~\ref{fig:fig2}, together with the dependence on the gas surface density predicted by the fiducial model of \citetalias{kruijssen12d} (solid line). It is clear that the observations by \citetalias{chandar15} provide a good match to the \citetalias{kruijssen12d} model when (1) the sample is restricted to the gravitationally bound clusters that are described by the model and (2) the comparison is based on the area-normalised quantity ($\Sigma$) that $\Gamma$ is predicted to correlate with. The reasonable agreement with the open symbols shows that correcting for disruption is desirable, but not essential.

In this Letter, we have presented of a set of guidelines for testing current cluster formation theories in a meaningful way and have demonstrated their validity. Of course, these guidelines apply specifically to tests of the \citetalias{kruijssen12d} theory, but they should also hold more generally. Independently of which particular theory for the formation of gravitationally bound clusters is considered, it remains key to omit unbound structure from the observational sample, to adopt area-normalised (or scale-independent) physical quantities, and to ensure similar spatial coverage of gas, star, and cluster formation tracers. We therefore propose that these guidelines should be used as a general reference frame for future work.

Our results are of particular relevance in view of the currently-ongoing, large surveys of cluster formation across a broad range of galactic environments (e.g.~LEGUS and Hi-PEEC, see \citealt{calzetti15,adamo16}). Over the coming years, these optical/UV surveys will be complemented with ALMA surveys of the molecular gas from which the stellar clusters in these galaxies are forming. The careful comparison of these observations to current theories will enable the first systematic census of cluster formation physics in the nearby Universe.

\section*{Acknowledgements}
We thank Angela Adamo, Mark Krumholz, S\o ren Larsen, and Esteban Silva-Villa for helpful comments. The anonymous referee is acknowledged for a swift and constructive report that improved this work. JMDK is funded by a Gliese Fellowship and NB is partially funded by a Royal Society University Research Fellowship and a European Research Council Consolidator Grant.

\bibliographystyle{mnras}
\bibliography{mybib}

\bsp

\label{lastpage}

\end{document}